\documentclass[twocolumn,twoside,amsmath,amssymb,superscriptaddress,showpacs,nofootinbib,aps]{revtex4-1}

\usepackage{slashed}
\usepackage{amsmath}
\usepackage{amsthm}
\usepackage[utf8]{inputenc}
\usepackage{graphicx}
\usepackage{epsfig}
\usepackage{amssymb}
\usepackage{dcolumn}
\usepackage{bm}
\usepackage{color}

\newcommand{\ba}{\begin{array}}
\newcommand{\ea}{\end{array}}
\def\br{\begin{eqnarray}}
\def\er{\end{eqnarray}}
\def\be{\begin{equation}}
\def\ee{\end{equation}}
\usepackage{hyperref}

\def\({\left(}
\def\){\right)}

\def\<{\left\langle}
\def\>{\right\rangle}

\newcommand{\Q}{\textnormal{\tiny \textsc{Q}}}
\newcommand{\T}{\textnormal{\tiny \textsc{T}}}
\newcommand{\E}{\textnormal{\tiny \textsc{E}}}
\def\tt{\textnormal\tiny\textsc}

\begin{document}

\title{Fermion mass splitting in the technicolor coupled scenario}

\author{A. Doff}
\email{agomes@utfpr.edu.br}

\affiliation{Universidade Tecnol\'ogica Federal do Paran\'a - UTFPR - DAFIS
Av Monteiro Lobato Km 04, 84016-210, Ponta Grossa, PR, Brazil}

\author{A. A. Natale} 
\email{adriano.natale@unesp.br}

\affiliation{Instituto de F{\'i}sica Te\'orica - UNESP, Rua Dr. Bento T. Ferraz, 271,\\ Bloco II, 01140-070, S\~ao Paulo, SP, Brazil}

\begin{abstract}
We discuss fermion mass generation in unified models where QCD and technicolor (or any two strongly interacting theories) have
their Schwinger-Dyson equations coupled. In this case the technicolor (TC) and QCD self-energies are modified in comparison with the
behavior observed in the isolated theories. In these models the pseudo-Goldstone boson masses are much higher than the ones obtained in different contexts, and phenomenological signals, except from a light scalar composite boson, will be quite difficult to be observed at present collider energies. The most noticeable
fact of these models is how the mass splitting between the different ordinary fermions is generated. We discuss how a necessary horizontal 
(or family) symmetry can be implemented in order to generate the mass splitting between fermions of different generations; how the fermionic mass spectrum may
be modified due to GUT interactions, as well as how the mass splitting within the same fermionic generation are generated due to electroweak and GUT interactions. 
\end{abstract}



\maketitle

\section{introduction}

The hierarchy and triviality problems related to the existence of fundamental scalar bosons have been discussed for a long time. The first
attempts to solve these problems were proposed forty years ago in the seminal papers by Weinberg~\cite{wei} and Susskind~\cite{sus}. In these
works the fundamental scalar boson that would be responsible for the Standard Model (SM) gauge symmetry breaking was substituted by a composite
scalar boson generated by a new strong interaction dubbed as Technicolor (TC). This proposal was incorporated in the model of Farhi 
and Susskind~\cite{fs} together with the idea that Nature may also have a Grand Unified Theory (GUT). These type of models were reviewed in 
Refs.~\cite{far,mira}. The possible existence of composite scalar bosons and a GUT are beautiful and naturally expected ideas. It is worth
remembering that much that it was learned up to now about symmetry breaking involves a composite scalar boson (as in the QCD chiral symmetry breaking
and in the microscopic BCS theory of superconductivity), and the SM convergence of interactions at high energy seems to indicate the presence of a
GUT. Unfortunately, it is also known how difficult is to build a phenomenologically viable model along these lines~\cite{hs,og1,og2}.

TC models continue to be studied although they present several phenomenological problems~\cite{hs,an,sa1,sa2,sa3,sa4,be}. Most of these problems are related to the soft behavior of the technifermions self-energy, and they can be ameliorated if the TC theory has a large mass anomalous dimension as proposed by
Holdom several years ago, what leads to a harder behavior for the TC self-energy~\cite{holdom}. Models following this idea, called as
walking technicolor and variations, started to be investigated, and this large anomalous dimension can be produced with
the introduction of a large number of fermions in a fundamental fermionic representation, as well as with fermions in larger dimensional representation, or with the introduction of an effective four-fermion interaction (a partial list of these works appear in Refs.~\cite{wa1,wa2,wa3,wa4,wa5,wa6,wa7,wa8,wa9,wa10,wa11,wa12,wa13,wa14,wa15,wa16,wa17,wa18,takeuchi}). 

In the models described above the necessary extended technicolor (ETC) boson masses, that usually induce
flavor changing neutral currents, can be pushed to higher energies; the pseudo-Goldstone boson masses are enhanced and the composite scalar boson mass, that plays the role of the Higgs boson, turns out
to be light, what does not happen if the TC theory has the usual dynamical behavior of an isolated non-Abelian theory.
It is interesting that recent lattice simulations of $SU(3)$ gauge theories with $N_f=8$ fundamental Dirac fermions~\cite{la1,la2,la3,la4,la5} and $N_f=2$ symmetric sextets Dirac
fermions~\cite{la6,la7,la8,la9,la10} show evidence of a light scalar boson, as predicted by TC theories with a self-energy behavior modified by a large mass anomalous dimension. 

All these approaches to solve the TC problems have the start point of an isolated strong interaction non-Abelian gauge theory, where what is known about QCD is modified
in order to verify how the TC self-energy changes with momentum. These attempts imply theories with a large number of technifermions or fermions in higher dimensional
representations. Here we discuss a new approach consisting of models where TC and QCD are coupled through a larger theory~\cite{us1,us2,us3}, leading naturally
to a theory with a light scalar boson, where the ETC (or unified theory) gauge boson masses can be pushed to very high energies and pseudo-Goldstone boson masses are enhanced. All this happens without the need of a strong increase of the number of technifermions or the introduction of larger dimensional fermionic representations. In the TC coupled
scenario the necessity of having a system located near a conformal fixed point, is not so pressing in order to have a light composite Higgs boson as it seems to happen in the many
lattice calculations-\cite{la1,la2,la3,la4,la5,la6,la7,la8,la9,la10,la11}.  

The most striking point of these models is that the hierarchy between different ordinary fermion masses do not appear as a consequence
of different ETC boson masses, but has its origin in the presence of a necessary horizontal (or family) symmetry. Moreover, the coupled strong interactions
also fit naturally with the idea of a larger or grand unified theory. Here the fermionic mass splitting is discussed in different contexts, showing how they may be generated and 
which are the advantages of the TC coupled scenario, where even the GUT existence has deep implications in the fermionic mass generation. In Section II we briefly review some aspects of the TC coupled scenario. In Section III we show how the mass splitting between different fermionic generations are intimately connected
to the presence of a horizontal symmetry, and how a GUT participate in the generation of the fermionic mass spectrum. In Section IV we discuss how electroweak interactions may affect the mass splitting within fermions of the same generation, as well as the existence of a GUT may also imply isodoublets mass differences.

\section{The TC coupled scenario}

We would like to recall the results of Ref.~\cite{us1,us2} to see
what happens in the chiral symmetry breaking of coupled strong interaction theories. In Ref.~\cite{us1} we calculated numerically the  self-energy ($\Sigma (p^2)$) of two coupled strong interaction theories (QCD and a $SU(2)$ TC theory), corresponding to the diagrams shown in the first line of Fig.(1) for the techniquarks ($T$) coupled to the quarks ($Q$) by
some ETC or GUT.
\begin{figure}[h]
\centering
\includegraphics[scale=0.5]{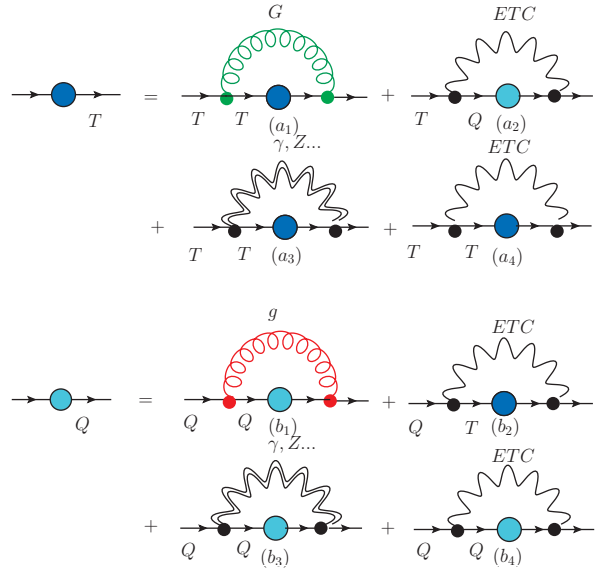} 
\vspace{-0.25cm} 
\caption[dummy0]{The coupled system of  SDEs for TC  ($T\equiv$technifermion) and QCD  ($Q\equiv$quark). This system may also include  ETC and electroweak or any other corrections, and some of these are indicated in the figure. $G \,(g)$ indicates a technigluon (gluon).}
\label{fig1}
\end{figure}

We verified numerically that the coupled TC self-energy behaves as
\be
\Sigma_{\T}(p^2)\approx \mu_{\tt{TC}} \left[ 1+ \delta_1 \ln\left[(p^2+\mu^2_{\tt{TC}})/\mu^2_{\tt{TC}}\right] \right]^{-\delta_2} \,,
\label{eq1}
\ee
where $\mu_{\tt{TC}}$ is the dynamical TC mass, which should be of the order of the Fermi scale. $\delta_1$ and $\delta_2$ are parameters that depend on the QCD, TC and ETC theory. In the case with more interactions (e.g. electroweak) these parameters will contain corrections proportional to the charges of these
theories. Note that Eq.(\ref{eq1}) is the simplest interpolation of the numerical result of Ref.~\cite{us1}, describing the infrared (IR) dynamical mass proportional
to $\mu_{\tt{TC}}$ (or $\mu_{\tt{QCD}}$), and a logarithmic decreasing function of the momentum in the ultraviolet (UV) region.

The behavior of Eq.(\ref{eq1}) is not a surprise. The fact that another interaction added to the TC one changes the self-energy is known since the work of
Takeuchi~\cite{takeuchi}. The reason for the behavior described above is that as TC give masses to ordinary fermions QCD also give masses to 
the technifermions~\cite{us2}, as well as other interactions may
contribute to these masses when all their Schwinger-Dyson equations (SDE) are coupled as shown in Fig.(1). In Ref.\cite{us2} we verified analytically that the second
diagram on the right-hand side of Fig.(1) for techniquarks and quarks modify the UV boundary condition of the SDE in differential form exactly as happens in
the case of bare massive fermions, i.e. Eq.(\ref{eq1}) looks similar to the self-energy of a fermion with a bare mass. However, the $\delta_2$ value which would be proportional to the mass anomalous
dimension of a technifermion, has now its value connected to the QCD dynamics (that generates the technifermion mass) as well as to the other interactions
present in the unified system. The quarks self-energies are also modified accordingly to Eq.(\ref{eq1}).

There is a very simple way to understand the result of Eq.(\ref{eq1}). Let us assume that the QCD and TC self-energies are coupled 
through an ETC (or GUT) interaction
with coupling constant $g_{ETC}$, all of them asymptotically free, and that the effect of each strong interaction (QCD or TC) is to proportionate an effective bare mass to each other. The UV asymptotic behavior of the self-energies will have the following form
\be
\Sigma_{UV} (p^2)\approx \mu \left( \frac{p^2}{\mu^2}\right)^{-\Delta (g^2,g_{ETC}^2)} \,,
\label{eq1b}
\ee
where $\mu$ and $g$ may represent QCD or TC dynamical masses and couplings, and $\Delta (g^2,g_{ETC}^2)$ is a function of the asymptotically small couplings. 
The self-energy UV
behavior is expected to be dominated by the ETC (or GUT) interaction, whereas the IR behavior of the self-energy is dominated by the strong interaction, although we
should not expect a simple expression for $\Delta (g^2,g_{ETC}^2)$. Expanding Eq.(\ref{eq1b}) for small couplings we obtain exactly an expression similar
to the one of Eq.(\ref{eq1}).

The ordinary fermion masses generated by Eq.(\ref{eq1}), as shown in Ref.~\cite{us1,us2,us3}, will be given by
\be
m_{\Q} \approx \lambda_E \mu_{\tt{TC}} [1+\kappa_1 \ln(M^2_{\E}/\mu_{\tt{TC}}^2)]^{-\kappa_2} \, ,
\label{eq2}
\ee
where $\lambda_E$ involves ETC couplings and a Casimir operator eigenvalue, $M_E$ is an ETC boson mass, and the $\kappa_i$ are also functions of the $\delta_i$  in 
Eq.(\ref{eq1}) as well as other possible corrections (electroweak or other interactions). Frequently we will approximate the ordinary fermion masses just
by 
\be
m_{\Q} \propto \lambda_E \mu_{\tt{TC}} \, , 
\label{eq2r}
\ee
where we are assuming that the contribution between brackets in Eq.(\ref{eq2}) is small. Note that self-energies with such logarithmic behavior (or similar integrals) have
a slowly convergent  behavior (see, for instance, Ref.\cite{brod} or the appendix of Ref.\cite{luna}), and have its result dominated by the factor
$\lambda_E \mu $ (where $\mu$ can be the TC or QCD dynamical mass).

In the coupled scenario the QCD self-energy \textit{has the same behavior of Eq.(\ref{eq1})},
only changing $\mu_{\tt{TC}}$ by $\mu_{\tt{QCD}}$ (the QCD dynamical mass) and respective $\delta_i$ coefficients. Therefore, the infrared (IR) behavior of TC and QCD self-energies are proportional to their respective strongly generated dynamical masses ($\mu_{\tt{TC}}$ and $\mu_{\tt{QCD}}$), while their UV behavior is the one 
of ``hard" dynamically generated masses, whose perturbative anomalous dimension is dominated by the different strong interaction than the one that determines their IR behavior. Some consequences of such behavior are discussed in the sequence.

\subsection{A light composite scalar boson}

The traditional wisdom about the composite scalar boson mass that plays the role of the Higgs boson is that it should be of the order of the Fermi scale. This
idea appeared at the first time in the seminal work of Nambu and Jona-Lasinio~\cite{nl}, where the scalar composite mass of the strong interaction (the $\sigma$ meson)was determined as
\be
m_\sigma = 2 \mu_{\tt{QCD}}\,\, .
\label{eq3}
\ee
In QCD the same result for the $\sigma$ meson (now known as $f_0(500)$~\cite{pdg}) was obtained in Ref.~\cite{ds} through the solution of the homogeneous Bethe-Salpeter equation (BSE).
In the TC case we should also have the same result, leading to an expected composite Higgs mass $m_H = 2 \mu_{\tt{TC}}$. Once $\mu_{\tt{TC}}$ is of order
of the Fermi scale, this mass would be heavy and of order of a TeV. This high mass value is surely not the case of the observed Higgs boson~\cite{atlas,cms}.

The result of Eq.(\ref{eq3}) is correct if QCD (or TC) are considered as isolated theories, which are characterized by a very soft self-energy, decreasing with the momentum
as $1/p^2$. When the self-energy is hard, like the one of Eq.(\ref{eq1}){\footnote{When solving the SDE for the fermion mass we have two possible solutions, one behaving as $\mu^3/p^2$ that is called 
regular or soft, and another behaving as $\mu(p^2/\mu^2)^{-\gamma}$ called irregular or hard, where $\gamma$ is proportional to $bg^2$ and $b$ is the leading
coefficient of the $\beta$ function. This irregular or hard 
solution is similar to an explicit chiral symmetry breaking~\cite{mi}, i.e. to the existence of a bare fermion mass, and this is exactly what happens in the TC 
coupled case, where the extra SDE diagrams indeed act to provide ``bare" masses to the fermions. 
The irregular expression expanded for small $bg^2$, {\it {whose limit also coincides with the nearly 
conformal case (i.e. small $b$ value)}} leads exactly to Eq.(\ref{eq1}) 
where $\delta_1=bg^2$, as demonstrated in the appendix of 
Ref.~\cite{cors}.}}, the scalar composite mass should be determined by the non-homogenous BSE, which correspond to the
homogenous equation constrained by a normalization condition. The importance of the normalization 
condition was established a long time ago by Mandelstam~\cite{man} 
(see also Ref.~\cite{chl}) and in the
QCD case it was discussed by Lane~\cite{lane2}. The effect of the BSE normalization condition is fundamental to reduce the scalar mass determination when the BSE
wave function decreases slowly with the momentum. The calculation of this effect, considering a self-energy given by Eq.(\ref{eq1}), has been performed in 
Ref.~\cite{dnp} and \textit{imply in a decrease of the scalar mass estimate by one order of magnitude} (see also Ref.~\cite{dln}). A similar analysis using an effective potential for composite
operators and a hard self-energy corroborates with this result~\cite{dnp2}.

\subsection{Ordinary fermion and ETC gauge boson masses}

The ordinary fermion masses are determined through the diagram (A) of Fig.(2), where an ETC gauge boson connects the different fermions and technifermions. The mass
splitting between different fermionic generations is usually thought as a consequence of different ETC gauge boson masses. It is clear from Eq.(\ref{eq2}) that
this is not the case in the coupled scenario. The ordinary fermion masses vary logarithmically with $M_E$, and as proposed in Refs.~\cite{us1,us3} the fermion mass
splitting is induced by a horizontal (or family) symmetry, where, at leading order, the third generation couples only to the TC condensate and the first generation 
to the QCD condensate. The second generation results from the mixing of the different condensates intermediated by the horizontal (or GUT) bosons~\cite{us3}. These
points are going to be discussed at lenght in the next sections.

The main result of the coupled scenario is that the different fermionic mass scales are a consequence of the different strong interactions. Another point, as
can be verified in Refs.~\cite{us3}, is that quarks are heavier than leptons due to the large number of diagrams contributing to their masses. The ETC
gauge boson masses can be pushed to very high energies, where the symmetry breaking of the ETC (or GUT) model can be generated even by fundamental scalar bosons, that
may appear naturally at Planck scale. Therefore, we do not expect large flavor changing neutral current problems associated to ETC. Finally, the fact that 
technifermions couple at leading order only to the third fermionic family will hide much of the technicolor signals involving light quarks or leptons (i.e.
first or second fermionic generation).  

\subsection{Pseudo-Goldstone boson masses}

Unified TC models contain a large number of Goldstone bosons that appear when the TC chiral symmetry is broken. Most of the technifermions obtain masses due to
radiative corrections leading to quite heavy pseudo-Goldstone bosons. As one example of how large are their masses in the TC coupled scenario, let us consider the lightest technifermion, which in many cases may be a techni-neutrino. This particle will appear when technifermions are a doublet of the weak interactions, they should be
there if we want to give masses to the charged leptons, and its mass appear due to the diagram (B) of Fig.(2).

\begin{figure}[t]
\centering
\includegraphics[width=1\columnwidth]{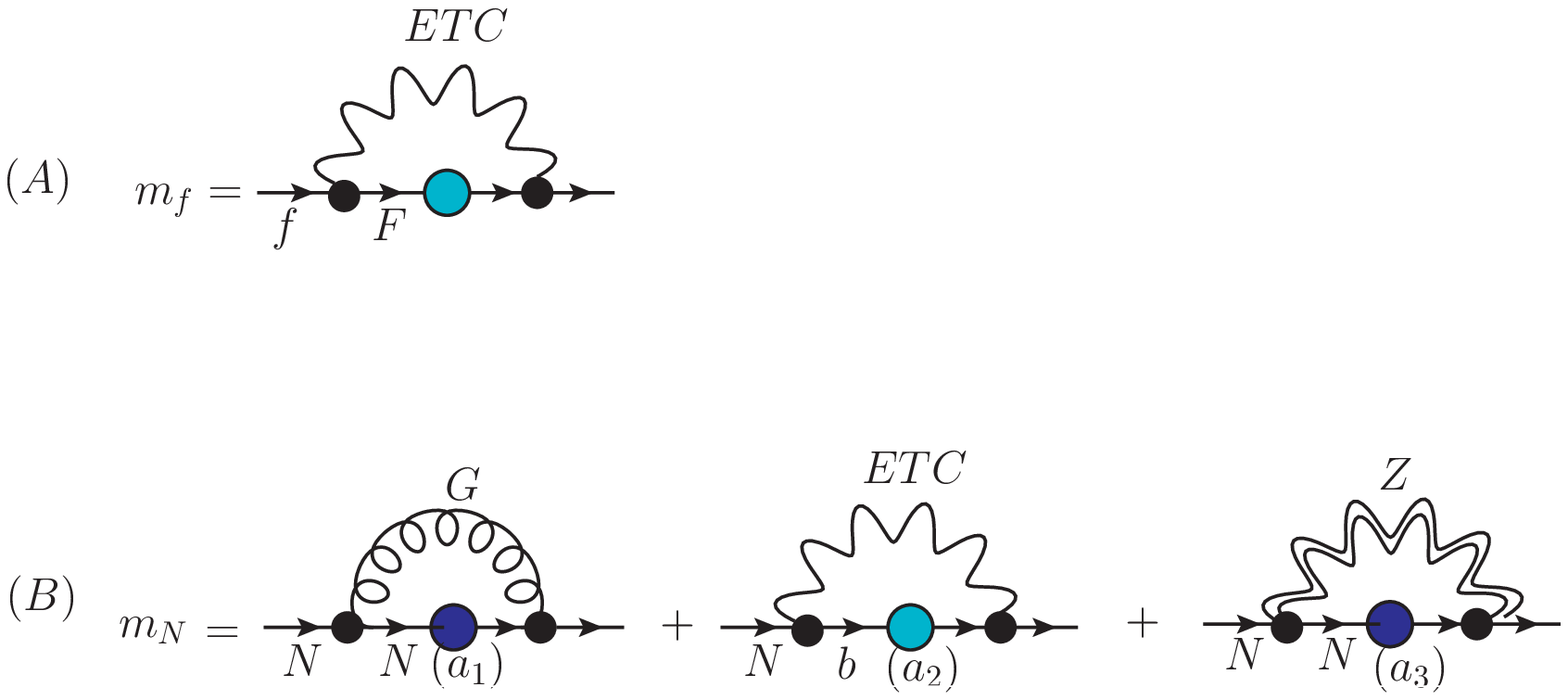}
\caption[dummy0]{The standard mass diagram in TC theories is shown in part (A) of the figure. It connects ordinary fermion to technifermions
through ETC interactions. In the coupled TC scheme more diagrams are taken into account, e.g. the techni-neutrino mass has all the contributions shown in part (B) of the figure. }
\label{fig2}
\end{figure}

The diagram ($a_2$) of Fig.(2) induces the logarithmic running of the techni-neutrino self-energy, and the third diagram of Fig.(2) provides a current
mass ($m_N$) of weak origin of order $g^2_w \mu_{\tt{TC}}\approx{\cal{O}}(100)$GeV. Simply using the Gell-Mann-Oakes-Renner relation 
$m^2_\Pi \approx m_N \frac{\< {\bar{N}}N\>}{2F_{\Pi}^2}$, with $\<{ \bar{N}}N\>\approx (250)^3$GeV$^3$ and $F_{\Pi}\approx 190$GeV (the technipion decay constant),
we roughly obtain for the lightest pseudo-Goldstone boson mass formed by techni-neutrinos a mass of the following order~\cite{us3}
\be
m_{\Pi}^N \approx 150 \,\,  GeV.
\label{mpn}
\ee
All other pseudo-Goldstone masses will be heavier than this one, and \textit{at leading order they couple only to the third fermionic generation what is enforced by
the horizontal symmetry}. Therefore, there is a great probability that they may have escaped detection up to now.

\section{Model building and fermion mass splittings for different generations}

A very simple fact of model building in the coupled scenario is that we do not need a TC group larger than $SU(2)$, which is enough to produce the
gauge symmetry breaking of the electroweak group. It was usually expected that the TC group should be larger than $SU(3)$ just in order to
condensate at one larger scale than QCD, assuming that the TC group emerges in a tumbling scheme~\cite{sus2}. As the generated masses depend logarithmically on the ETC (or GUT) interaction, these theories can be pushed to very high energies, and in this case it is possible to assume that the larger group symmetry, that contains QCD and TC, can be broken by fundamental scalars, which
may appear naturally at GUT or Planck scales (perhaps due to the presence of supersymmetry). Therefore, a simple $SU(2)$ TC theory can appear at the
Fermi scale in this breaking .

As technifermions and ordinary fermions will interact among themselves in the coupled scenario, the left-handed techniquarks will transform as $(3,2)$ 
SM representations, i.e. color triplets and electroweak doublets, right-handed techniquarks as $(3,1)$, left-handed technileptons as $(1,2)$ and right-handed
technileptons as $(1,1)$. The number of technifermions ($N_{TF}$) will be
\be
N_{TF}=3 N_Q + N_L \,\, ,
\label{eq4}
\ee
where $N_Q$ is the number of techniquarks and $N_L$ the number of technileptons. Therefore, the smallest number of technifermions will be $8$, which, if the
TC group is $SU(2)$, is a number that implies that we are already in the walking window~\cite{holdom,wa1,wa2,wi1,wi2,wi3}.

The hard self-energy behavior of the strong interactions (TC and QCD) represented in Eq.(\ref{eq1}) works in the direction of smaller contributions to the
$S$ parameter~\cite{us3}. Choosing a TC group as small as $SU(2)$ and technifermions in a small representation of
dimension $d(R_{TC})$ it is also interesting. If we have a small number of TC doublets $(N_D)$, a $SU(2)_{TC}$ group and just assuming a naive expression for the $S$ parameter
\be
S=N_D \frac{d(R_{TC})}{6\pi} \,\, .
\label{eq4b}
\ee
we may probably obtain this parameter within the expected experimental limits.

In the coupled case the hierarchy of ordinary fermion masses will not appear as a result of different ETC gauge boson masses (see Eq.(\ref{eq2}) and the discussion in
Refs.~\cite{us1,us3}), but is originated due to the presence of a discrete or continuous horizontal (or family) symmetry. For simplicity we can consider a 
simple Lie group $G$
\be
G\supset G_H \times G_U \,\, ,
\label{eq5}
\ee 
where $G_H$ is the horizontal group symmetry and $G_U$ contains
\be
G_U \supset G_{ETC}\times SU(2)_L \times U(1)_Y  \,\, ,
\label{eq5b}
\ee
where 
\be
G_{ETC} \supset SU(N)_{TC} \times SU(3)_c \,\, ,
\label{eq5bb}
\ee
contains all strongly interacting theories, as in the model of Ref.~\cite{fs}.
Another possibility is that
\be
G_U \supset SU(N)_{TC} \times G_{SM} \,\,
\label{eq5bc}
\ee
where $G_{SM}$ is the SM group
\be
G_{SM}=SU(3)_c \times SU(2)_L \times U(1)_Y \,\, .
\label{eq5c}
\ee
These possibilities were studied in Refs.~\cite{og1,og2}.

\par A more ambitious case appears when $G_U$ is an unified group\,\, containing\,\, the\,\, TC\,\, group\,\, and the Georgi-Glashow 
\\ $SU(5)_{GUT}$~\cite{gg}. $G_U$ can be a $SU(N)$ group such that
\be
G_U \equiv SU(N)_U \supset SU(N_{TC})\times SU(5)_{GUT} \,\, .
\label{eq6}
\ee
It follows that $N$ should be equal to $N_{TC}+5$ with the minimal value $N=7$, although this choice contains only two ordinary fermion generations~\cite{fs}.

Here we shall not worry about the symmetry breaking of the large unified groups, since we assume that this breaking always can be promoted by fundamental
scalars once it can occur at very high energies. The type of the horizontal symmetry (be it
global or gauge, discrete or continuous) will also
not be discussed as long as a set of anomaly free representations is found for the $G$ group. As the ETC (or GUT) gauge boson masses will be very heavy
we will not expect the presence of flavor changing interactions at one undesirable level, and we will be mostly concerned with the origin of the fermionic
mass splittings, which is the most interesting characteristic of this type of model, where the mass difference between different fermionic generations
is related to the different strong interactions present in the theory.

In the scheme represented in Eq.(\ref{eq6}), the case where $SU(N)_U \equiv SU(7)$ is not realistic since it contains only two generations, but it can be 
studied because is a nice example of how the different $SU(7)$ and $SU(5)_S$\footnote{As in Ref.~\cite{fs}, we assume the symmetry breaking direction $SU(7) \to [SU(2)_{L}\times U(1)_Y]\times SU(5)_{S}$, where the strong $SU(5)_{S}$ gauge theory acts as a extended technicolor theory (ETC)  and as considered in Ref.~\cite{fs} we shall not discuss the $SU(7)$(or $SU(5)_{S}$) symmetry breaking.} interactions generate the mass splitting between the different 
fermionic generations. In this case we can assume that the $SU(7)$ gauge symmetry breaking is produced by fundamental scalars bosons at a very high unification
scale, and the same happens with the $SU(5)_S$ strong gauge group that breaks into QCD and TC (although this last breaking could be a result of tumbling~\cite{sus2}).
We consider the following set of anomaly free $SU(7)$ antisymmetric representations
\\
\br
&&\left[2\right]=(1,10)+(2,5)+(1,1), \nonumber \\
&&\left[4\right]=(1,10)+(2,{\bar{10}})+(1,{\bar{5}}), \nonumber\\
&&\left[6\right]=(1,{\bar{5}})+(2,1) \,\, ,
\label{eqc7}
\er
\\
as in Ref.~\cite{fs}, but choosing the first and third ordinary fermionic representations in order to show how QCD and TC act in the generation
of these fermion mass scales. We differ from Ref.~\cite{fs} in the fact that the $[2]$ representation now is
\\
\br 
&&(1,10)=\left(\begin{array}{ccccc} 0 & \bar{t}_r & -\bar{t}_y & t_r & b_r  \\  -\bar{t}_r & 0 & \bar{t}_b & t_y & b_y  \\ \bar{t}_y & -\bar{t}_b & 0 & t_b & b_b \\ -t_r & -t_y& -t_b & 0 & \bar{\tau} \\ -b_r & -b_y & -b_b & -\bar{\tau} & 0 \end{array}\right) \nonumber \\
&&(2,5)=\left(\begin{array}{c} D_r \\ D_y\\ D_b \\ \bar{E} \\ \bar{N} \end{array}\right)_p\,\,,\,\, (1,1) = \bar{\nu}_{\tau} 
\label{eqc8}
\er
\\
\noindent where technifermions are represented by capital letters and the $[4]$ is decomposed as
\\
\br 
&&(1,10)=\left(\begin{array}{ccccc} 0 & \bar{u}_r & -\bar{u}_y & u_r & d_r  \\  -\bar{u}_r & 0 & \bar{u}_b & u_y & d_y  \\ \bar{u}_y & -\bar{u}_b & 0 & u_b & d_b \\ -u_r & -u_y & -u_b & 0 & \bar{e} \\ -d_r & -d_y & -d_b  & -\bar{e} & 0 \end{array}\right) \,\,,\,\,(1,\bar{5}) = \left(\begin{array}{c} \bar{b}_r \\ \bar{b}_y\\ \bar{b}_b \\ \tau \\ \nu_{\tau} \end{array}\right)\nonumber \\ 
&&(2,\bar{10}) = \left(\begin{array}{ccccc} 0 & U_r & -U_y & \bar{U}_r & \bar{D}_r \\  -U_r & 0 & U_b & \bar{U}_y & \bar{D}_y  \\ U_y & -U_b & 0 & \bar{U}_b & \bar{D}_b \\ -\bar{U}_r & -\bar{U}_y& -\bar{U}_b & 0 & E \\ -\bar{D}_r & -\bar{D}_y & -\bar{D}_b & -E & 0 \end{array}\right)_p.  
\label{eqc9}
\er
\\
The representation $[6]$ is the same as the one in Ref.~\cite{fs} just exchanging the second by the third fermionic family.

Note the particular choice of Eqs.(\ref{eqc8}) and (\ref{eqc9}). This example is very nice because in this case we naturally couple only the third generation fermions to TC while the first
generation is coupled to QCD as shown in Fig.(3). 
\begin{figure}[t]
\centering
\includegraphics[width=0.8\columnwidth]{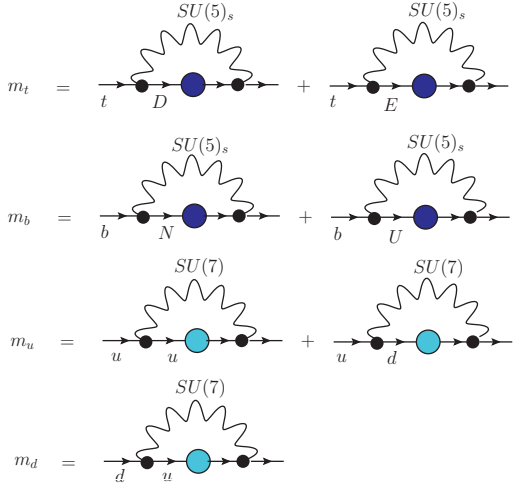}
\caption[dummy0]{ Quark masses in the coupled TC scheme of Eq.(\ref{eq5b}), for the fermionic content given by Eqs.(\ref{eqc8}) and (\ref{eqc9}). }
\label{fig3}
\end{figure}
However we still need mixing between the different generations and interactions, and this is provided by the introduction of an $SU(2)_H$ horizontal interaction that may
also be broken at the $SU(7)$ or $SU(5)_S$ scale. Note that $SU(2)_H$ is the simplest choice to obtain the cancellation of
anomalies (i.e., $ 2A[\bar{10} + 10] = 0$ and $2A[\bar{5} + 5]=0$), and it will generate the diagrams of Fig.(4).
\begin{figure}[h]
\centering
\includegraphics[width=0.8\columnwidth]{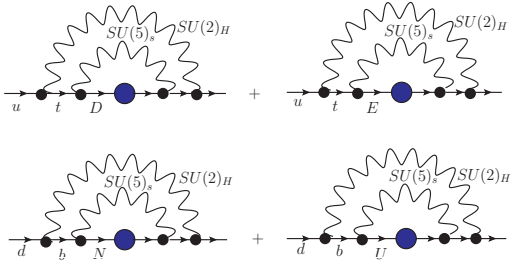}
\caption[dummy0]{Horizontal symmetry contribution to the b quark mass.}
\label{figx}
\end{figure}

Looking at Eq.(\ref{eq2}) and neglecting the logarithmic term between brackets we can assume that the generated fermion masses ($m_f$) are proportional to
\be
m_f \approx \lambda_i \mu_s \,\, ,
\label{eq41}
\ee
where
\be
\lambda_i = \alpha_i C_i \,\, ,
\label{eq41c}
\ee 
and where $\alpha_i$ is the coupling constant appearing in the mass diagram, with one fermion in a group representation that has a Casimir eigenvalue $C_i$, involving a
strong interaction with dynamical mass $\mu_s$ (which can be the QCD or TC dynamical mass). Therefore the full mass matrix for charge $2/3$ quarks in the $SU(2)_H$ 
(or ($u$ $t$) charge $2/3$ quarks) basis is due to the 
diagrams of Figs.(3) and (4) is
giving by
\\
\br 
m_{2/3} = \left(\begin{array}{cc} a  &  b\\  -b &  c \end{array}\right),
\label{eqm23}
\er
\\
\noindent where we can identify
\\
\br
&& a \approx \mu_{QCD} (C_H\alpha_H + 2C_7\alpha_7)\\
&& c \approx \mu_{TC} C_5\alpha_5  \\ 
&& b \approx \mu_{TC} C_5C_H\alpha_5\alpha_H.
\er 
\\
The indices $H$, $7$ and $5$ are respectively related to couplings and Casimir operators of the groups $SU(2)_H$,  $SU(7)$ and $SU(5)_S$. 
Assuming $C_5\alpha_5 = O(0.1)$, $\frac{C_H}{C_5} = \frac{3}{4}.\frac{10}{24} \approx \frac{1}{3}$, we can
estimate $C_H\alpha_H \approx \frac{O(0.1)}{3} $, and with naive round values of $\mu_{QCD} = 0.2 GeV$ and $\mu_{TC}= 1 TeV$ we obtain
\\
\br 
m_{2/3} = \left(\begin{array}{cc} 0.047  &  3\\  -3 &  100 \end{array}\right) \,\, ,
\er
\\
\noindent which, when diagonalized, gives
\br 
&& m_t \approx 100 GeV \\
&& m_u  \approx 0.1 GeV. 
\er
With similar mass values for the $1/3$ electrically charged quarks. These approximations are very rough but they provide a clear idea how
the mass splitting of the different fermionic families appear as a consequence of the different strong interactions present in the theory.

It is interesting to observe how lepton masses are generated when models are building according to the scheme of Eq.(\ref{eq5b}) or (\ref{eq6}), in these
cases the contributions of the larger or unified theories are still important, due to the logarithmic dependence on the masses of these interactions. For instance, in the model just described above the electron and tau lepton obtain masses through the diagrams of Fig.(5). We can again see that
the charged lepton of the first generation obtain mass coupling to the QCD condensate, while the third generation lepton obtain mass
coupling directly to the technifermion condensate, and the smallness of the leptonic masses compared to quark masses appear naturally, also
because leptons have less interaction with the strong interactions. 
\begin{figure}[t]
\centering
\includegraphics[width=1\columnwidth]{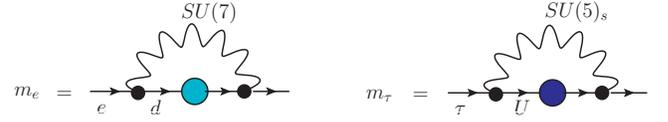}
\caption[dummy0]{ Leptonic masses in the TC coupled scenario generated in models of the type of Eq.(\ref{eq6}).}
\label{fig1}
\end{figure}

The discussion presented above shows how the existence of a GUT is important to the generation of
the fermionic mass spectrum, while the horizontal symmetry is fundamental to determine  the splitting of 
fermion masses. The choice of the horizontal symmetry is
tied to the existence of two fundamental strong interaction condensates: the TC and QCD condensates. This system acts as a coupled system of
two fundamental scalar bosons with two vacuum expectation values of the order of QCD and TC scales. Let us make the following association
\br
&& \< q \bar{q} \>\leftrightarrow \< \phi\> \\
&& \< Q \bar{Q} \> \leftrightarrow \< \Phi \> \, ,
\label{cond}
\er
where $\< q \bar{q} \>$ and $\< Q \bar{Q} \>$ are respectively the quark and techniquark condensates, which have been associated to composite
scalar fields $\phi$ and $\Phi$, with vacuum expectation values of approximately $250$ MeV and $250$ GeV. The fermionic horizontal quantum numbers 
for the first and third generation must be chosen such that the QCD condensate (or the composite $\phi$) couples only to the first generation.
While the quantum numbers of the third generation fermions are chosen in order to couple only to the TC condensate (or the composite $\Phi$).

In a realistic scenario, containing three generations of fermions, the horizontal symmetry (perhaps also the GUT interaction) will be responsible for the mixing of the composite fields, contributing to generate the masses of the second fermionic generation. Examples of horizontal symmetries are the simple $SU(2)_H$ case discussed above and the $SU(3)_H$ group discussed in Ref.~\cite{us3}. However, we envisage several other possibilities because the gauge bosons of this new interaction, or 
even the new fermions necessary to render the theory anomaly free, can be made very heavy once the symmetry breaking of the horizontal symmetry may happens at very high energies, due to the weak dependence of the fermion masses on the horizontal (or ETC and GUT) scales as shown in Eq.(\ref{eq2}).

When attempting to build a  
concrete and realistic model along the lines discussed here, we have to keep in mind that all physical parameters remain the same as in standard TC 
models, except that ETC (GUT) mass scales can be pushed to high energies, and it is necessary the introduction of a family (or horizontal)
symmetry in order to generate the mass splitting between different fermionic generations. Concerning the gauge group structure of TC coupled models
it is fundamental that QCD and TC are embedded into a larger group, generating a system of two composite scalar bosons responsible for the gauge and
chiral symmetry breaking. We have already proposed a model based on the $SU(9)$ group plus a family symmetry that has several of the
expected properties for a realistic model~\cite{us3}. A larger group is still possible
and even semi-simple groups are allowed as long as the strong interactions are coupled. We have also noted that it is possible to choose different
family symmetries (local or global). However, we have to say that it is quite difficult to obtain a realistic determination of all fermion masses.
What happens is that in the coupled models we must first solve a large set of coupled Schwinger-Dyson equations in order to determine
realistic TC and QCD self-energies, which is a hard task even computing SDE with rough approximations. In a first approach it is possible to neglect 
the most feeble interactions in the full SDE set of equations, although these other interactions are fundamental to generate the full mass spectra.
Further development along this line is necessary before expanding the space of possible theories, as the determination of a 
simple method to capture the relevant behavior of the coupled equations, otherwise it will be difficult to know how proposed models
are indeed realistic.

\section{Mass splitting within the same fermionic generation}

\subsection{Electroweak mass splitting}

In Ref.~\cite{us2} we transformed the ESD coupled system into a differential equation. Applying the boundary conditions to the resulting equation
it is possible to verify that the UV asymptotic behavior of the coupled system is given by
\be
\Sigma (p^2\rightarrow \infty )\propto a \mu \left( \frac{p^2}{\mu^2} \right)^{-\Delta(\kappa,\epsilon)} \,\, ,
\label{eq31}
\ee
where  $\Delta(\kappa,\epsilon) \approx \frac{\epsilon}{2} - \frac{\kappa}{8}$ assuming asymptotic linear superposition of the different interactions, in the expression above 
$a$ is a constant and $\mu$ can be either the TC or QCD dynamical mass, $\kappa$ is proportional do the ETC (or GUT) coupling constant and in the limit when this  coupling is zero  
\be
\Delta(\kappa,\epsilon)= \frac{\epsilon}{2} = \frac{3C\alpha}{4\pi} \,\, ,
\label{eq32}
\ee
where $C$ and $\alpha=g^2/4\pi$ are respectively the Casimir operator eigenvalue and coupling constant of the strong interaction dominating the self-energy solution of the coupling system (TC or QCD).

In the UV region we assume that Eq.(\ref{eq31}) may be expanded as function of the coupling constant of the strong  interaction, therefore  assuming $\Delta(\kappa,\epsilon) \propto b g^2$ we may perform the following expansion
\be
\Sigma (p^2\rightarrow \infty )\propto \mu \left[1+bg^2 ln\left(\frac{p^2}{\mu^2}\right)\right]^{-\frac{\Delta(\kappa,\epsilon)}{bg^2}} \, ,
\label{eq33}
\ee
which is one expression consistent with Eq.(\ref{eq1}). Assuming that we are in the deep UV region this last expression can be reduced to
\be
\Sigma_{UV} (p^2) \propto \mu 
\left( bg^2\right)^{-\frac{\Delta(\kappa,\epsilon)}{bg^2}}\left[ln\left(\frac{p^2}{\mu^2}\right) \right]^{-\frac{\Delta(\kappa,\epsilon)}{bg^2}} \, ,
\label{eq34}
\ee
which has exactly the form $A [ln (p^2/\mu^2)]^{-B}$ that we used in Ref.~\cite{us1} to fit the numerical solution of the coupled SDE system.
The above expressions will allow us to obtain a simple evaluation of how the splitting of fermion masses within the same generation may appear.

In Fig.(6) we show a simplified coupled system where two fermions, indicated by $1=U$ and $2=D$ obtain masses. 
\begin{figure}[t]
\begin{center}
\epsfig{file=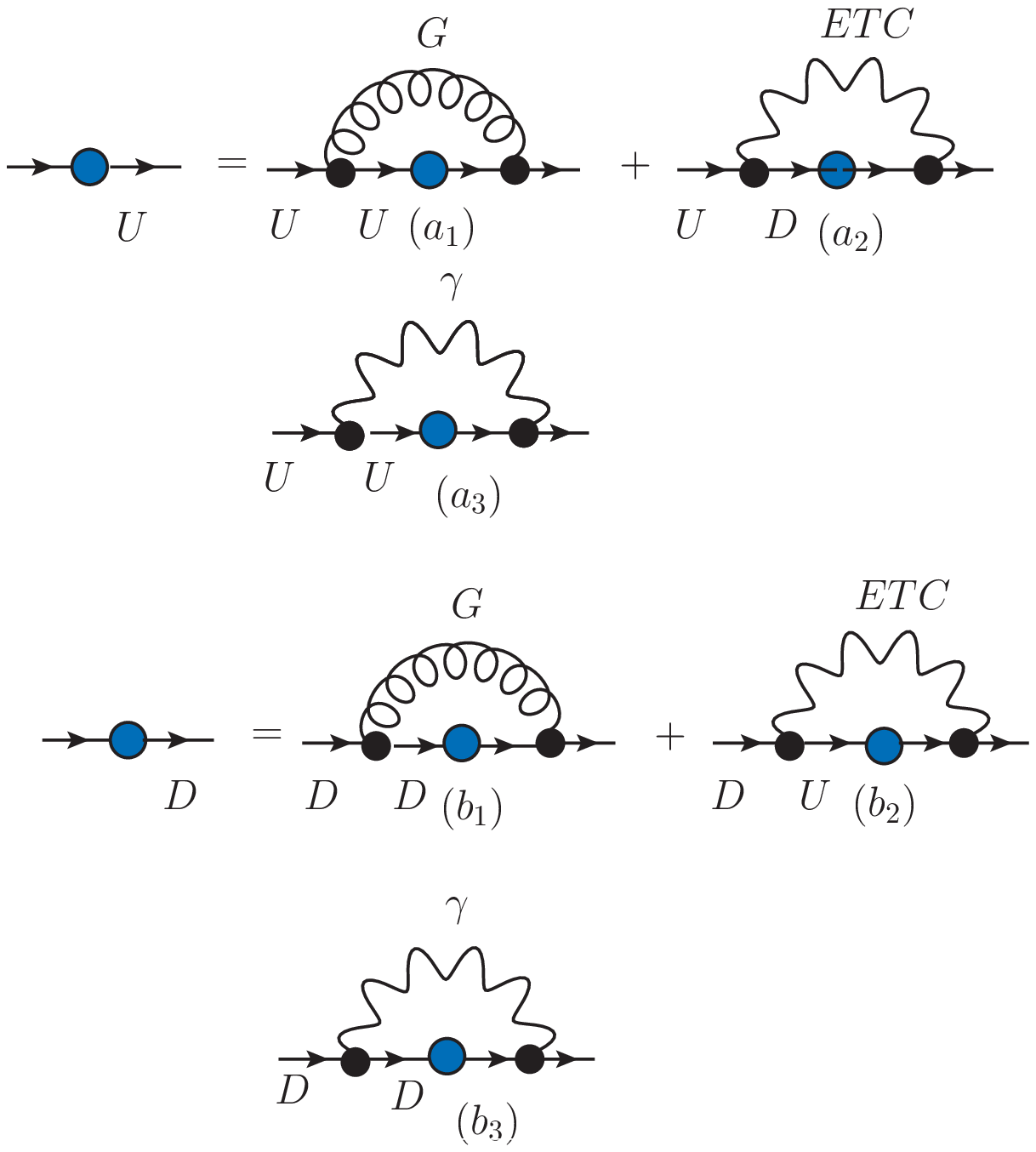,width=0.45\textwidth}
\caption{Coupled SDE system of two fermions, indicated by $1=U$ and $2=D$, considering also the electromagnetic interaction (3rd diagram).}
\end{center}
\end{figure} 
Besides the ETC and TC diagrams generating masses for the fermions $U$ and $D$ in Fig.(6), we also include the diagrams involving the electromagnetic interaction
($a_3, b_3$). If we consider only the first two contributions in Fig.(6) $(a_1, a_2, b_1, b_2)$, we can assume for the solution of
these diagrams that the variable $\Delta(\kappa,\epsilon)$ of Eqs.(\ref{eq33}) or (\ref{eq34}) can be approximated by
\be 
\Delta_E(\kappa,\epsilon)\equiv \Delta_1(\kappa,\epsilon) = \Delta_2(\kappa,\epsilon) = \frac{\epsilon}{2} - \frac{\kappa}{8}  
\label{eq48}
\ee
\noindent in such a way that the masses of the fermions $U$ and $D$, respectively indicated by $M_1$ and $M_2$ will have the following ratio
\be 
\frac{M_1}{M_2} = \left(\ln\left(\frac{M^2_{E}}{\mu^2}\right)\right)^{\frac{\frac{\epsilon}{2} - \frac{\kappa}{8}  - (\frac{\epsilon}{2} - \frac{\kappa}{8} ) }{bg^2}} = 1 \,\, .
\label{eq49}
\ee
\noindent in the above equation the index $E$ means that we considered the coupled system only with the ETC contribution, i.e. the system is coupled only due to the ETC interaction (diagrams $a_2$ and $b_2$).

If the diagrams containing the electroweak interaction are included we could expect a modification of the variables $\Delta_1(\kappa,\epsilon) $ and $\Delta_2(\kappa,\epsilon) $ in the following form   
\br
&& \Delta_1(\kappa,\epsilon) = \Delta_E(\kappa,\epsilon) + \delta^{1}_{\gamma}(\kappa)\nonumber \\
&& \Delta_2(\kappa,\epsilon)  = \Delta_E(\kappa,\epsilon) + \delta^{2}_{\gamma}(\kappa).
\label{eq50}
\er
\noindent Assuming that the electric charge of the fermions $U$ and $D$ differ by a factor of $2$, i.e. $|Q_1| = 2|Q_2|$, and it is the square of the charge that enter in these coefficients we can write
\br
&& \Delta_1(\kappa,\epsilon) = \Delta_E(\kappa,\epsilon) + 4\delta^{2}_{\gamma}(\kappa)\nonumber \\
&& \Delta_2(\kappa,\epsilon)  = \Delta_E(\kappa,\epsilon) + \delta^{2}_{\gamma}(\kappa) \,\, ,
\label{eq51}
\er
\noindent we can now see that the mass ratio will now behave as
\be 
\frac{M_1}{M_2} \approx \left(\ln\left(\frac{M^2_{E}}{\mu^2}\right)\right)^{\frac{3\delta^{2}_{\gamma}(\kappa)}{bg^2}} >  1 \, .
\label{eq52}
\ee
\noindent Note that this ratio is basically dependent on the UV behavior of the generated mass. The IR behavior is dominated by the strong interaction and is identical for both fermions and dominated by the diagrams $(a_1, b_1)$ of Fig.(6). The mass splitting comes from $ETC$(GUT) , $\gamma, W, Z...$ interactions that modify the parameters $\Delta_i(\kappa,\epsilon)$, although it is important to see that the factor $bg^2$ coming from the strong  interaction also controls the scale of the masses and intervene in the splitting.

Considering Eq.(\ref{eq52}) we can make a naive estimate: Suppose we have as TC group $SU(3)_{TC}$ with $n_f = 6$. Assuming also the MAC hypothesis we can approximate $bg^2 = 0.44$, that $M_E$ is at the GUT scale,
$\alpha  \approx \alpha_E= 0.032$ and $C_E=1$ we obtain $\delta^{2}_{\gamma}(\kappa) \sim 0.032$, with 
\be 
\frac{M_1}{M_2} \approx \left(\ln\left(\frac{M^2_{E}}{\mu^2}\right)\right)^{0.22} \,\, .
\label{eq53}
\ee
\noindent With $M_{E} = \Lambda_{ETC} = \Lambda_{GUT} = 10^{16} GeV = 10^{13} TeV $ and $\mu = 1TeV$ we obtain
\be 
\frac{M_1}{M_2}  = \left(\ln\left(\frac{M^2_{E}}{\mu^2}\right)\right)^{0.22} \approx 2.5.
\label{eq54}
\ee
The values of couplings and other constants used in this evaluation are the same ones used in our numerical solution of the coupled SDE in Ref.~\cite{us1}.

Note that mass splittings of an order of magnitude can also be obtained, particularly if we consider unified models like the ones proposed in the 
beginning of the section which may originate several diagrams contributing to the mass of one specific fermion. It is also clear that in this
very simple estimate the number of fermions also enters into account. For example, if we consider the walking limit for the $SU(3)_{TC}$ theory, and
instead of $n_f = 6$ we use $n_f = 12$ the $bg^2$ estimate in Eq.(\ref{eq52})
will be modified and a larger mass splitting than the one of Eq.(\ref{eq54}) can be obtained.

\subsection{GUT mass splitting}

Mass splitting for fermions within the same generation can be obtained, without large weak isospin violation, if we put the chiral components (LH and RH) of a given fermion member of an electroweak doublet in different ETC representations~\cite{si1,si2}.
For instance, let us consider how a large mass difference between the top and bottom quark can be generated. We assume the following group structure
\be
\hspace{2cm}{G} = SU(3)_{{}_{ETC}} \times G'
\ee
where $G'= SU(3)_{{}_{C}}
\times SU(2)_{{}_{L}}\times SU(2)_{{}_{R}} \otimes U(1)_{{}_{S}}$. We also consider that the gauge group $G'$ is embedded in  a $SO(12)$ gauge group \cite{raj}. Under the ETC interaction the $G$ group fermionic content can be decomposed as:
\br
&& \psi_{{}_{{{\bf{3}}}_{ETC}}} = \left(\begin{array}{c}   T_{1} \\ T_{2} \\ t \end{array}\right)_{L,R}\,\,,\,\, \left(\begin{array}{c}   B_{1} \\ B_{2} \\ b \end{array}\right)_{L}\,\,\sim \,\,({\bf{3}}, 1)
 \\ \nonumber \\
&& \psi_{{}_{\bar{\bf{3}}_{ETC}}} = \left(\begin{array}{c}  B_{1} \\ B_{2} \\ b \end{array}\right)_{R}\,\,\sim \,\,({\bar{\bf{3}}}, 1)
\er 
\noindent  where we are putting the LH and RH chiral components of the top quark in the same ETC representation, but these components for the botton quark are in different ETC representations. Observing the Feynman rules for this theory we can verify that the bottom
quark couples to the top quark at one loop through the intermediation of a heavy $SO(12)$ gauge boson, and also at two loops due to the interaction
with right and left-handed weak gauge bosons. The bottom mass turns out to be 
\be
m_b \propto \lambda_{12} m_t \, ,
\ee
where the diagram that generates this mass is similar to  the last diagram shown in Fig.(3), with the exchange u$\to$t,  d$\to$b and $SU(7)\to SO(12)$. The coefficient  $\lambda_{12}$ is proportional to the coupling and other constants of the $SO(12)$ theory, and the mass difference between these quarks can certainly be of one order of magnitude or larger.

The above example show to us how a GUT plays an 
important role in the mass splitting of ordinary fermions in the coupled scenario, since any very large gauge boson mass dependence does not affect
strongly the value of the generated mass. However we cannot forget that the necessary horizontal interactions 
may also induce a mass splitting for fermions in the same generation, which will depend on the particularities of the chosen horizontal group.

\section{Conclusions}

In this work we discuss unified models based on the TC coupled scenario. In these models the QCD and TC Schwinger-Dyson equations are coupled due
to ETC or GUT interactions. One theory provides mass to the other and their self-energies vary logarithmically with the momentum. In this case
the generated fermion masses are weakly dependent on the ETC or GUT gauge boson masses, and the fermionic mass splitting is not generated due to these
different gauge boson masses. Some of these points are briefly reviewed in Section II. The fermionic masses will be generated due to the presence of a horizontal (or family) symmetry, and this is the
most noticeable characteristic of these models. 

In Section III we discussed a very simple unified model with only two generations. The model is a variation of the model of Ref.~\cite{fs} where
by construction the first fermionic generation receives mass coupling only to QCD condensates, while the third generation couples only to the TC condensate.
Even in this model we verify that a horizontal symmetry must be introduced in order to reproduce the mixing between families. The existence of a
GUT also increase the number of diagrams contributing to one specific mass.

A series of possible unified models along this line have also been discussed. In the coupled scenario the QCD and TC are responsible for the 
hierarchy of fermion masses, i.e. this hierarchy is intimately connected to the different scales of strongly interacting theories. There shoud be a
variety of horizontal symmetry groups that can lead to an approximate estimate of the known fermion mass spectrum, but more specifically
the horizontal quantum numbers of fermions and technifermions must be chosen such that QCD and TC condensates (or composite scalars)
couple respectively to the first and third fermionic generation. Horizontal (ETC or GUT) mass scales can be pushed to high energies due
to the small dependence of the fermion masses on these scales, and the choice of anomaly free representations of the horizontal symmetry
should not hamper model building.

In Section IV we show how an estimate of the mass splitting within the same fermionic generation can be obtained. A very simple estimate of electroweak
effects is presented, where it appears without the introduction of many fermions (or walking behavior). However, it is also possible to see that
the walking limit certainly increases the inter-generation mass splitting. We also discuss how a GUT can also induce a large mass splitting for the same iso-doublet. 

The few aspects discussed here show that models along the line of coupled strong interactions may open way for the construction of realistic theories of dynamical symmetry breaking.

\section*{Acknowledgments}
This research  was  partially supported by the Conselho Nacional de Desenvolvimento Cient\'{\i}fico e Tecnol\'ogico (CNPq)
under the grants 302663/2016-9 (A.D.) and 303588/2018-7 (A.A.N.).

\end{document}